\documentclass{sig-alternate-arxiv}

\usepackage{graphicx}
\usepackage{stfloats}

\begin{document}
%
\conferenceinfo{ACM Multimedia Workshop on Affect \& Sentiment}{'15 Brisbane, Australia}

\title{Diving Deep into Sentiment: Understanding Fine-tuned CNNs for Visual Sentiment Prediction}


\numberofauthors{1} 
\author{
\alignauthor V{\'i}ctor Campos\raisebox{9pt}{$\ast$} \,\,\, Amaia Salvador\raisebox{9pt}{$\ast$} \,\,\, Brendan Jou\raisebox{9pt}{$\dagger$} \,\,\, Xavier Gir{\'o}-i-Nieto\raisebox{9pt}{$\ast$}\\
       \affaddr{\raisebox{9pt}{$\ast$}Universitat Polit{\`e}cnica de Catalunya (UPC), Barcelona, Catalonia/Spain}\\
       \affaddr{\raisebox{9pt}{$\dagger$}Columbia University, New York, NY USA}\\
       \email{victor.campos.camunez@alu-etsetb.upc.edu, \{amaia.salvador,xavier.giro\}@upc.edu, bjou@ee.columbia.edu}
}

\maketitle
\begin{abstract}
Visual media are powerful means of expressing emotions and sentiments. The constant generation of new content in social networks highlights the need of automated visual sentiment analysis tools. While Convolutional Neural Networks (CNNs) have established a new state-of-the-art in several vision problems, their application to the task of sentiment analysis is mostly unexplored and there are few studies regarding how to design CNNs for this purpose. In this work, we study the suitability of fine-tuning a CNN for visual sentiment prediction as well as explore performance boosting techniques within this deep learning setting. Finally, we provide a deep-dive analysis into a benchmark, state-of-the-art network architecture to gain insight about how to design patterns for CNNs on the task of visual sentiment prediction.
\end{abstract}




\section{Introduction}
\begin{figure}
	\includegraphics[width=\linewidth]{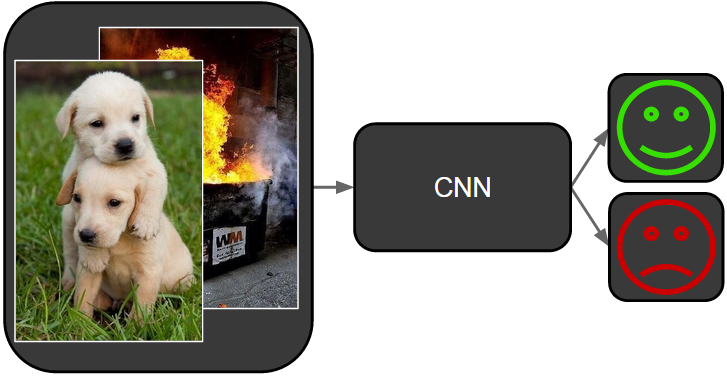}
    \caption{Overview of the presented system for visual sentiment prediction.}
		\label{fig:introduction}
\end{figure}

The recent growth of social networks has led to an explosion in amount, throughput and variety of multimedia content generated every day.
One reason for the richness of this social multimedia content comes from how it has become one of the principal ways that users share their feelings and opinions about nearly every sphere of their lives.
In particular, visual media, like images and videos, have risen as one of the most pervasively used and shared documents in which emotions and sentiments are expressed.

The advantages of having machines capable of understanding human feelings are numerous and would imply a revolution in fields such as robotics, medicine or entertainment.
Some interesting preliminary applications are already beginning to emerge, e.g.~for emotional understanding of viewer responses to advertisements using facial expressions \cite{mcduffpredicting}.
However, while machines are approaching human performance on several recognition tasks, such as image classification \cite{he_2015}, the task of automatically detecting sentiments and emotions from images and videos still presents many unsolved challenges.
Numerous approaches towards bridging the \emph{affective gap}, or the conceptual and computational divide between low-level features and high-level affective semantics, have been presented over the years for visual multimedia \cite{machajdik_2010,jia_2012,borth_2013,jou_2014}, but the performance has remained fairly conservative and related intuitions behind this have been lacking.

Promising results obtained using Convolutional Neural Networks (CNNs) \cite{lecun_1998} in many fundamental vision tasks have led us to consider the efficacy of such machinery for higher abstraction tasks like sentiment analysis, i.e.~classifying the visual sentiment (either positive or negative) that an image provokes to a human.
Recently, some works \cite{you_2015,xu2014visual} explored CNNs for the task of visual sentiment analysis and obtained some encouraging results that outperform the state of the art, but develop very little intuition and analysis into the CNN architectures they used.
%
%
Our work focuses on acquiring insight into fine-tuned layer-wise performance of CNNs in the visual sentiment prediction setting.
We address the task of assessing the contribution of individuals layers in a state-of-the-art fine-tuned CNN architecture for visual sentiment prediction.



Our contributions include:
(1) a visual sentiment prediction framework that outperforms the state-of-the-art approach on an image dataset collected from Twitter using a fine-tuned CNN,
(2) a rigorous analysis of layer-wise performance in the task of visual sentiment prediction by training individual classifiers on feature maps from each layer in the former CNN, and
(3) network architecture surgery applied to a fine-tuned CNN for visual sentiment prediction.

\section{Related Work}
Several approaches towards overcoming the gap between visual features and affective semantic concepts can be found in the literature. In \cite{siersdorfer_2010}, the authors explore the potential of two low-level descriptors common in object recognition, Color Histograms (LCH, GCH) and SIFT-based Bag-of-Words, for the task of visual sentiment prediction. Some other works have considered the use of descriptors inspired by art and psychology to address tasks such as visual emotion classification \cite{machajdik_2010} or automatic image adjustment towards a certain emotional reaction \cite{peng_2015}. In \cite{borth_2013} a Visual Sentiment Ontology based on psychology theories and web mining consisting of 3,000 Adjective Noun Pairs (ANP) is built. These ANPs serve as a mid-level representation that attempt to bridge the affective gap, but they are very dependent on the data that was used to build the ontology and are not completely suitable for domain transfer.

The increase in computational power in GPUs and the creation of large image datasets such as \cite{deng_2009} have allowed Deep Convolutional Neural Networks (CNNs) to show outstanding performance in computer vision challenges \cite{krizhevsky_2012, szegedy_2014_googlenet, he_2015}. And despite requiring huge amounts of training samples to tune their millions of parameters, CNNs have proved to be very effective in domain transfer experiments \cite{oquab_2014}. 
This interesting property of CNNs is applied to the task of visual sentiment prediction in \cite{xu2014visual}, where the winning architecture of ILSVRC 2012 \cite{krizhevsky_2012} (5 convolutional and 3 fully connected layers) is used as a high-level attribute descriptor in order to train a sentiment classifier based on Logistic Regression. Although the authors do not explore the possibility of fine-tuning, they show how the off-the-shelf CNN descriptors outperform hand-crafted low-level features and SentiBank \cite{borth_2013}. Given the distinct nature of visual sentiment analysis and object recognition, the authors in \cite{you_2015} explore the possibility of designing a new architecture specific for the former task, training a network with 2 convolutional and 4 fully connected layers. However, there is very little rationale given for why they configured their network in this way except for the last two fully connected layers. 
Our work focuses on fine-tuning a CNN for the task of visual sentiment prediction and later performing a rigorous analysis of its architecture, in order to shed some light on the problem of CNN architecture designing for visual sentiment analysis.

\section{Methodology}
The Convolutional Neural Network architecture employed in our experiments is \emph{CaffeNet}, a slight modification of the ILSVRC 2012 winning architecture, \emph{AlexNet} \cite{krizhevsky_2012}. This network, which was originally designed and trained for the task of object recognition, is composed by 5 convolutional layers and 3 fully connected layers. The two first convolutional layers are followed by pooling and normalization layers, while a pooling layer is placed between the last convolutional layer and the first fully connected one. The experiments were performed using \emph{Caffe} \cite{jia_2014}, a publicly available deep learning framework.

We adapted \emph{CaffeNet} to a sentiment prediction task using the Twitter dataset collected and published in \cite{you_2015}. This dataset contains 1,269 images labeled into positive or negative by 5 different annotators. 
The choice was made based on the fact that images in Twitter dataset are labeled by human annotators, oppositely to other annotation methods which rely on textual tags or predefined concepts. Therefore, the Twitter dataset is less noisy and allows the models to learn stronger concepts related to the sentiment that an image provokes to a human.
Given the subjective nature of sentiment, different subsets can be formed depending on the number of annotators that agreed on their decision. Only images that built consensus among all the annotators (5-agree subset) were considered in our experiments. 
The resulting dataset is formed by 880 images (580 positive, 301 negative), which was later divided in 5 different folds to evaluate experiments using cross-validation.

Each of the following subsections is self-contained and describes a different set of experiments. Although the training conditions for all the experiments were defined as similar as possible for the sake of comparison, there might be slight differences given each individual experimental setup. For this reason, every section contains the experiment description and its training conditions as well.

\subsection{Fine-tuning CaffeNet}
The adopted \emph{CaffeNet} \cite{jia_2014} architecture contains more than 60 million parameters, a figure too high for training the network from scratch with the limited amount of data available in the Twitter dataset. Given the good results achieved by previous works about transfer learning \cite{oquab_2014, salvador_2015}, we decided to explore the possibility of fine-tuning an already existing model. Fine-tuning consists in initializing the weights in each layer except the last one with those values learned from another model. The last layer is replaced by a new one, usually containing the same number of units as classes in the dataset, and randomly initializing their weights before ``resuming'' training but with inputs from the target dataset. The advantage of this approach compared to fully re-training a network from a random initialization on all the network weights is that it essentially starts the gradient descent learning from a point much closer to an optimum, reducing both the number of iterations needed before convergence and decreasing the likelihood of overfitting when the target dataset is small.

In our sentiment analysis task, the last layer from the original architecture, \emph{fc8}, is replaced by a new one composed of 2 neurons, one for \emph{positive} and another for \emph{negative} sentiment. The model of \emph{CaffeNet} trained using ILSVRC 2012 dataset is used to initialize the rest of parameters in the network for the fine-tuning experiment. Results are evaluated using 5-fold cross-validation. They are all fine-tuned during 65 epochs (that is, every training image was seen 65 times by the CNN), with an initial base learning rate of 0.001 that is divided by 10 every 6 epochs. As the weights in the last layer are the only ones which are randomly initialized, its learning rate is set to be 10 times higher than the base learning rate in order to provide a faster convergence rate.

A common practice when working with CNNs is data augmentation, consisting of generating different versions of an image by applying simple transformations such as flips and crops. Recent work has proved that this technique reports a consistent improvement in accuracy \cite{chatfield_2014}. We explored whether data augmentation improves the spatial generalization capability of our analysis by feeding 10 different combination of flips and crops of the original image to the network in the test stage. The classification scores obtained for each combination are fused with an averaging operation.

\subsection{Layer by layer analysis}

\begin{figure}
		\includegraphics[width=\linewidth]{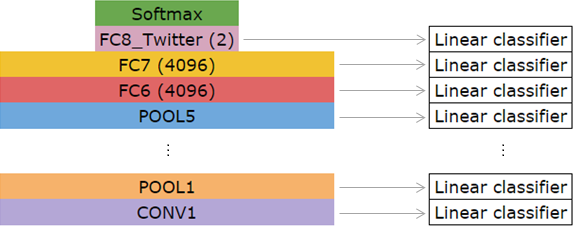}
		\caption{Experimental setup for the layer analysis using linear classifiers. The number between brackets next to fully connected layer makes reference to the amount of neurons they contain.}
		\label{fig:linear_classfiers}
\end{figure}

Despite the outstanding performance of CNNs in many vision tasks, there is still little intuition into how to design them. In order to gain some insight about the contribution of each individual layer to the the task of visual sentiment prediction, we performed an exhaustive layer-per-layer analysis of the fine-tuned network.

The outputs of individual layers 
have been previously used as visual descriptors \cite{razavian_2014, salvador_2015}, where each neuron's activation is seen as a component of the feature vector. Traditionally, top layers have been selected for this purpose \cite{xu2014visual} as they are thought to encode high-level information. We further explore this possibility by using each layer as a feature extractor and training individual classifiers for each layer's features (see Figure \ref{fig:linear_classfiers}). This study allows measuring the difference in accuracy between layers and gives intuition not only about how the overall depth of the network might affect its performance, but also about the role of each type of layer, i.e. CONV, POOL, NORM and FC, and their suitability for visual sentiment prediction.

Neural activations in fully connected layers can be represented as \emph{d}-dimensional vectors, being \emph{d} the amount of neurons in the layer, so no further manipulation is needed. This is not the case of earlier layers, i.e. CONV, NORM, and POOL, whose feature maps are multidimensional, e.g. feature maps from \emph{conv5} are ${256x13x13}$ dimensional. These feature maps were flattened into \emph{d}-dimensional vectors before using them for classification purposes.
Two different linear classifiers are considered: Support Vector Machine with linear kernel and Softmax. The same 5-fold cross-validation procedure followed in the previous experiment is employed, training independent classifiers for each layer. Each classifier's regularization parameter is optimized by cross-validation.

\newpage  
\subsection{Layer ablation}
\label{section:methodology-layer_removal}

\begin{figure*}
		\includegraphics[width=\textwidth]{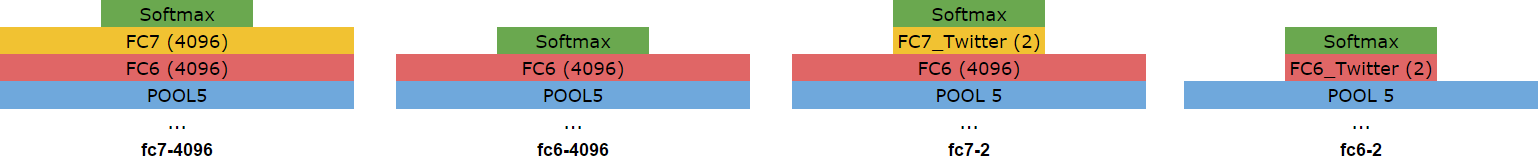}
		\caption{Layer ablation architectures. Networks \emph{fc7-4096} and \emph{fc6-4096} keep the original configuration after ablating the layers in the top of the architecture (Section \ref{layer_removal_1}), while in \emph{fc7-2} and \emph{fc6-2} the last remaining layer is replaced by a 2-neuron layer (as described in Section \ref{layer_removal_2}). The number between brackets next to fully connected layer makes reference to the amount of neurons they contain.} 
		\label{fig:layer_removal_architectures}
\end{figure*}

More intuition about the individual contribution of each layer can be gained by modifying the original architecture prior to training. This task is addressed by fine-tuning altered versions of the original \emph{CaffeNet} where top layers had been successively removed.

Different approaches to the layer removal problem might be taken, depending on the changes made to the remaining architecture. In our experiments, two different strategies are adopted: (1) a raw ablation by keeping the original configuration and weights for the remaining layers, and (2) adding a 2-neuron layer as a replacement to the removed one, on top of the remaining architecture and just before the Softmax layer. A more detailed definition of the experimental setup for each configuration is described in the following subsections.

\subsubsection{Raw ablation} 
\label{layer_removal_1}
In this set of experiments, the Softmax layer is placed on top of the remaining architecture, e.g. if \emph{fc8} and \emph{fc7} are removed, the output of \emph{fc6} is connected to the input of the Softmax layer. For the remaining layers, weights from the original model are kept as well.

The configurations studied in our experiments include versions of \emph{CaffeNet} where (1) \emph{fc8} has been ablated, and (2) both \emph{fc8} and \emph{fc7} have been removed (architectures \textit{fc7-4096} and \textit{fc6-4096}, respectively, in Figure \ref{fig:layer_removal_architectures}).
The models are trained during 65 epochs, with a base learning rate of 0.001 that is divided by 10 every 6 epochs. With this configuration all the weights are initialized using the pre-trained model, so random initialization of parameters is not necessary. Given this fact, there is no need to increase the individual learning rate of any layer.

\subsubsection{2-neuron on top}
\label{layer_removal_2}
As described in Section 3.1, fine-tuning consists in replacing the last layer in a net by a new one and use the weights in a pre-trained model as initialization for the rest of layers. Inspired by this procedure, we decided to combine the former methodology with the layer removal experiments: instead of leaving the whole remaining architecture unmodified after a layer is removed, its last remaining layer is replaced by a 2-neuron layer with random initialization of the weights.

This set of experiments comprises the fine-tuning of modified versions of \emph{CaffeNet} where (1) \emph{fc8} has been removed and \emph{fc7} has been replaced by a 2-neuron layer, and (2) \emph{fc8} and \emph{fc7} have been ablated and \emph{fc6} has been replaced by a 2-neuron layer (architectures \textit{fc7-2} and \textit{fc6-2}, respectively, in Figure \ref{fig:layer_removal_architectures}). The models are trained during 65 epochs, dividing the base learning rate by 10 every 6 epochs and with a learning rate 10 times higher than the base one for the 2-neuron layer, as its weights are being randomly initialized. The base learning rate of the former configuration is 0.001, while the latter's was set to 0.0001 to avoid divergence.

\subsection{Layer addition}

\begin{figure}[b]
		\includegraphics[width=\linewidth]{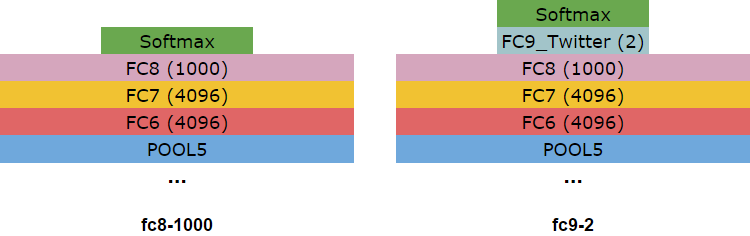}
		\caption{Architectures using the information contained in the original \emph{fc8} layer and weights. Configuration \emph{fc8-1000} reuses the whole architecture and weights from \emph{CaffeNet}, while \emph{fc9-2} features an additional 2-neuron layer. The number between brackets next to fully connected layer makes reference to the amount of neurons they contain.}
		\label{fig:using_fc8_information}
\end{figure}

None of the architectures that have been introduced so far takes into account the information encoded in the last layer (\emph{fc8}) of the original \emph{CaffeNet} model. This layer contains a confidence value for the image belonging to each one of the 1,000 classes in ILSVRC 2012. In addition, fully connected layers contain, by far, most of the parameters in a Deep Convolutional Neural Network. Therefore, from both of the former points of view, a remarkable amount of information is being lost when discarding the original \emph{fc8} layer in \emph{CaffeNet}. 

Similarly to the procedure followed in the layer removal experiments, two different approaches are considered in order to take advantage of the information in the original \emph{fc8}: (1) the original \emph{CaffeNet} architecture is fine-tuned, keeping the original configuration and weights for \emph{fc8}, and (2) a 2-neuron layer (\emph{fc9}) is added on top of the original architecture (architectures \emph{fc8-1000} and \emph{fc9-2}, respectively, in Figure \ref{fig:using_fc8_information}). Models are trained during 65 epochs, with a base learning rate of 0.001 that is divided by 10 every 6 epochs. The only layer that has a higher individual learning rate is the new \emph{fc9} in configuration \emph{fc9-2}, which is set to be 10 times higher than the base learning rate, given that its weights are randomly initialized.

\section{Experimental Results}
This section presents the results for the experiments proposed in the previous section, as well as intuition and conclusions.

\subsection{Fine-tuning CaffeNet}
\label{section:fine-tuning_results}

\begin{table} 
	\centering
	\caption{5-fold cross-validation results on 5-agree Twitter dataset}
    \label{fine-tuning_results}
    \resizebox{\linewidth}{!}{
      \begin{tabular}{|c|c|}
        \hline
        Model&Accuracy\\
        \hline
        Fine-tuned CNN from You et al. \cite{you_2015}&0.783 \\
        \hline
        Fine-tuned \emph{CaffeNet}&0.817 $\pm$ 0.038 \\
        \hline
        Fine-tuned \emph{CaffeNet} with oversampling&\textbf{0.830 $\pm$ 0.034 }\\
        \hline
      \end{tabular}
    }
\end{table}

Average accuracy results over the 5 folds for the fine-tuning experiment are presented in Table \ref{fine-tuning_results}, which also includes the results for the best fine-tuned model in \cite{you_2015}. This CNN, with a 2CONV-4FC architecture, was designed specifically for visual sentiment prediction and trained using almost half million sentiment annotated images from Flickr dataset \cite{borth_2013}. The network was finally fine-tuned on the Twitter 5-agree dataset with a resulting accuracy of 0.783 which is, to best of our knowledge, the best result on this dataset so far.

Surprisingly, fine-tuning a net that was originally trained for object recognition reported higher accuracy in visual sentiment prediction than a CNN that was specifically trained for that task. On one hand, this fact suggests the importance of high-level representations such as semantics in visual sentiment prediction, as transferring learning from object recognition to sentiment analysis actually produces high accuracy rates. On the other hand, it seems that visual sentiment prediction architectures also benefit from a higher amount of convolutional layers, as suggested by \cite{zeiler_2014} for the task of object recognition.

Averaging the prediction over modified versions of the input image results in a consistent improvement in the prediction accuracy. This behavior, which was already observed by the authors of \cite{chatfield_2014} when addressing the task of object recognition, suggests that the former procedure also increases the network's generalization capability for visual sentiment analysis, as the final prediction is far less dependent on the spatial distribution of the input image. 

\subsection{Layer by layer analysis}
\label{linear_classifiers_results}

\begin{table} 
	\centering
	\caption{Layer analysis with linear classifiers: 5-fold cross-validation results on 5-agree Twitter dataset}
    \label{table:linear_classifiers_results}
	\begin{tabular}{|c|c|c|}
      \hline
      Layer&SVM&Softmax\\
      \hline
      \emph{fc8} & 0.82 $\pm$ 0.055 & 0.821 $\pm$ 0.046 \\
      \hline
      \emph{fc7} & 0.814 $\pm$ 0.040 & 0.814 $\pm$ 0.044 \\
      \hline
      \emph{fc6} & 0.804 $\pm$ 0.031 & 0.81 $\pm$ 0.038 \\
      \hline
      \emph{pool5} & 0.784 $\pm$ 0.020 & 0.786 $\pm$ 0.022 \\
      \hline
      \emph{conv5} & 0.776 $\pm$ 0.025 & 0.779 $\pm$ 0.034 \\
      \hline
      \emph{conv4} & 0.794 $\pm$ 0.026 & 0.781 $\pm$ 0.020 \\
      \hline
      \emph{conv3} & 0.752 $\pm$ 0.033 & 0.748 $\pm$ 0.029 \\
      \hline
      \emph{norm2} & 0.735 $\pm$ 0.025 & 0.737 $\pm$ 0.021 \\
      \hline
      \emph{pool2} & 0.732 $\pm$ 0.019 & 0.729 $\pm$ 0.022 \\
      \hline
      \emph{conv2} & 0.735 $\pm$ 0.019 & 0.738 $\pm$ 0.030 \\
      \hline
      \emph{norm1} & 0.706 $\pm$ 0.032 & 0.712 $\pm$ 0.031 \\
      \hline
      \emph{pool1} & 0.674 $\pm$ 0.045 & 0.68 $\pm$ 0.035 \\
      \hline
      \emph{conv1} & 0.667 $\pm$ 0.049 & 0.67 $\pm$ 0.032 \\
      \hline
    \end{tabular}
\end{table}


The results of the layer-by-layer analysis of the fine-tuned \emph{CaffeNet} are presented in Table \ref{table:linear_classifiers_results}, both for the SVM and Softmax classifiers.

Recent works have studied the suitability of Support Vector Machines for classification using deep learning descriptors \cite{razavian_2014} while others have also replaced the Softmax loss by a SVM cost in the network architecture \cite{tang_2013}. Given the results of our layer-wise analysis, it is not possible to claim that any of the two classifiers provides a consistent gain compared to the other for visual sentiment analysis, at least in the Twitter 5-agree dataset with the proposed network architecture.

Accuracy trends at each layer reveal that the depth of the networks contributes to the increase of performance. Not every single layer produces an increase in accuracy with respect to the previous one, but even in those stages it is hard to claim that the architecture should be modified as higher layers might be benefiting from its effect, e.g. \emph{conv5} and \emph{pool5} report lower accuracy rates than earlier \emph{conv4} when their feature maps are used for classification, but later fully connected layers might be benefiting from the effect of \emph{conv5} and \emph{pool5} as all of them report higher accuracy than \emph{conv4}. 

An increase in performance is observed with each fully connected layer, as every stage introduces some gain with respect to the previous one. This fact suggests that adding additional fully connected layers might report even higher accuracy rates, but further research is necessary to evaluate this hypothesis.

\subsection{Layer ablation}
\label{section:layer_ablation_results}


\begin{table} 
	\centering
	\caption{Layer ablation: 5-fold cross-validation results on 5-agree Twitter dataset.
    }
    \label{table:layer_removal_results}
	\resizebox{\linewidth}{!}{
      \begin{tabular}{|c|c|c|}
        \hline
        Architecture&Without oversampling&With oversampling\\
        \hline
        fc7-4096 & 0.759 $\pm$ 0.023 & 0.786 $\pm$ 0.019\\
        \hline
        fc6-4096 & 0.657 $\pm$ 0.040 & 0.657 $\pm$ 0.040\\
        \hline
        fc7-2 & 0.784 $\pm$ 0.024 & 0.797 $\pm$ 0.021\\
        \hline
        fc6-2 & 0.651 $\pm$ 0.044 & 0.676 $\pm$ 0.029\\
        \hline
      \end{tabular}
	} 
\end{table}

The four ablation architectures depicted in Figure \ref{fig:layer_removal_architectures} are compared in Table \ref{table:layer_removal_results}.
These results indicate that
replacing the last remaining layer by a 2-neuron fully connected layer is a better solution than reusing the information of existing layers from a much higher dimensionality.
One reason for this behavior might be the amount of parameters in each architecture, as replacing the last layer by one with just 2 neurons produces a huge decrease in the parameters to optimize and, given the reduced amount of available training samples, that reduction can become beneficial.

Accuracy is considerably reduced when ablating \emph{fc7} and setting \emph{fc6} to be the last layer, independently of the method that was used. Further research revealed that models learned for architecture \textit{fc6-4096} always predict towards the majority class, i.e. positive sentiment, which is justified by the reduced amount of training data. This behavior is not observed in architecture \textit{fc6-2}, where the amount of parameters is highly reduced in comparison to \textit{fc6-4096}, but its performance is still very poor. Nevertheless, this result is somehow expected, as the convergence from a vector dimensionality 9,216 in \emph{pool5} to a layer with just 2 neurons might be too sudden. These observations suggest that a single fully connected layer might not be useful for the addressed task.


Finally, it is important to notice that networks which are fine-tuned after ablating \emph{fc8}, i.e. architectures \textit{fc7-4096} and \textit{fc7-2}, provide accuracy rates which are very close to the fine-tuned CNN in \cite{you_2015} or even higher. These results, as shown by the authors in \cite{zeiler_2014} for the task of object recognition, suggest that removing one of the fully connected layers (and with it, a high percentage of the parameters in the architecture) only produces a slight deterioration in performance, but the huge decrease in the parameters to optimize might allow the use of smaller datasets without overfitting the model. This is a very interesting result for visual sentiment prediction given the difficulty of obtaining reliable annotated images for such task.

\subsection{Layer addition}

\begin{table} 
	\centering
	\caption{Layer addition: 5-fold cross-validation results on 5-agree Twitter dataset.}
    \label{table:using_fc8_information_results}
    \resizebox{\linewidth}{!}{
      \begin{tabular}{|c|c|c|}
      
      	\hline
        Architecture&Without oversampling&With oversampling\\
        \hline
        fc8-1000 & 0.723 $\pm$ 0.041 & 0.731 $\pm$ 0.036\\
        \hline
        fc9-2 & 0.795 $\pm$ 0.023 & 0.803 $\pm$ 0.034 \\
        \hline
      
      \end{tabular}
    } 
\end{table}

The architectures that keep \emph{fc8} are evaluated in Table \ref{table:using_fc8_information_results}, indicating that architecture \textit{fc9-2} outperforms \textit{fc8-1000}. This observation, together with the previous in Section \ref{section:layer_ablation_results}, strengthens the thesis that CNNs deliver a higher performance in classification tasks when the last layer contains one neuron for each class.


The best accuracy results when reusing information from the original \emph{fc8} are obtained by adding a new layer, \emph{fc9}, although they are slightly worse than those obtained with the regular fine-tuning (Table \ref{fine-tuning_results}). At first sight, this observation may seem contrary to intuition gained in the layer-wise analysis, which suggested that a deeper architecture would have a better performance. If a holistic view is taken and not only the network architecture is considered, 
we observe that including information from the 1,000 classes in ILSVRC 2012 (e.g. zebra, library, red wine) may not help in sentiment prediction, as they are mainly neutral or do not provide any sentimental cues without contextual information. 


The reduction in performance when introducing semantic concepts that are neutral with respect to sentiment, together with the results in Section \ref{linear_classifiers_results}, highlight the importance of appropriate mid-level representation such as the Visual Sentiment Ontology built in \cite{borth_2013} when addressing the task of visual sentiment prediction. Nevertheless, they suggest that generic features such as neural codes in \emph{fc7} outperform semantic representations when the latter are not sentiment specific. 
This intuition meets the results in \cite{xu2014visual}, where the authors found out that training a classifier using \emph{CaffeNet}'s \emph{fc7} instead of \emph{fc8} reported better performance for the task of visual sentiment prediction.

\section{Conclusions}
We presented several experiments studying the suitability of fine-tuned CNNs for the task of visual sentiment prediction. 
We showed the utility of deep architectures that are capable of capturing high level features when addressing the task, obtaining models that outperform the best results so far in the evaluation dataset. 
Data augmentation has been demonstrated to be a useful technique for increasing visual sentiment prediction accuracy as well.
Our study of domain transfer from object recognition to sentiment analysis has reinforced common good practices in the field: discarding the last fully connected layer adapted to another task, and the addition of a new randomly initialized layer with as many neurons as the amount of categories to classify.



\section{Acknowledgments}
This work has been developed in the framework of the project BigGraph TEC2013-43935-R, funded by the Spanish Ministerio de Econom\'ia y Competitividad and the European Regional Development Fund (ERDF). 
The Image Processing Group at the UPC is a SGR14 Consolidated Research Group recognized and sponsored by the Catalan Government (Generalitat de Catalunya) through its  AGAUR office.
We gratefully acknowledge the support of NVIDIA Corporation with the donation of the GeForce GTX Titan Z used in this work.



\nocite{*} 
\bibliographystyle{abbrv}
\bibliography{sigproc} 

\begin{thebibliography}{10}

\bibitem{borth_2013}
D.~Borth, R.~Ji, T.~Chen, T.~Breuel, and S.-F. Chang.
\newblock Large-scale visual sentiment ontology and detectors using adjective
  noun pairs.
\newblock In {\em ACM MM}, 2013.

\bibitem{chatfield_2014}
K.~Chatfield, K.~Simonyan, A.~Vedaldi, and A.~Zisserman.
\newblock Return of the devil in the details: Delving deep into convolutional
  nets.
\newblock In {\em British Machine Vision Conference}, 2014.

\bibitem{deng_2009}
J.~Deng, W.~Dong, R.~Socher, L.-J. Li, K.~Li, and L.~Fei-Fei.
\newblock Image{N}et: {A} large-scale hierarchical image database.
\newblock In {\em Computer Vision and Pattern Recognition, 2009. CVPR 2009.
  IEEE Conference on}, pages 248--255. IEEE, 2009.

\bibitem{he_2015}
K.~He, X.~Zhang, S.~Ren, and J.~Sun.
\newblock Delving deep into rectifiers: Surpassing human-level performance on
  {I}mage{N}et classification.
\newblock {\em arXiv:abs/1502.01852 [cs.CV]}, 2015.

\bibitem{jia_2012}
J.~Jia, S.~Wu, X.~Wang, P.~Hu, L.~Cai, and J.~Tang.
\newblock Can we understand van {G}ogh's mood?: {L}earning to infer affects
  from images in social networks.
\newblock In {\em ACM MM}, 2012.

\bibitem{jia_2014}
Y.~Jia, E.~Shelhamer, J.~Donahue, S.~Karayev, J.~Long, R.~Girshick,
  S.~Guadarrama, and T.~Darrell.
\newblock Caffe: {C}onvolutional architecture for fast feature embedding.
\newblock In {\em ACM MM}, 2014.

\bibitem{jiang_2014}
Y.-G. Jiang, B.~Xu, and X.~Xue.
\newblock Predicting emotions in user-generated videos.
\newblock In {\em AAAI}, 2014.

\bibitem{jin_2010}
X.~Jin, A.~Gallagher, L.~Cao, J.~Luo, and J.~Han.
\newblock The wisdom of social multimedia: {U}sing {F}lickr for prediction and
  forecast.
\newblock In {\em ACM MM}, 2010.

\bibitem{jou_2014}
B.~Jou, S.~Bhattacharya, and S.-F. Chang.
\newblock Predicting viewer perceived emotions in animated {GIF}s.
\newblock In {\em ACM MM}, 2014.

\bibitem{kim_2013}
Y.~Kim, H.~Lee, and E.~M. Provost.
\newblock Deep learning for robust feature generation in audiovisual emotion
  recognition.
\newblock In {\em ICASSP}, 2013.

\bibitem{krizhevsky_2012}
A.~Krizhevsky, I.~Sutskever, and G.~E. Hinton.
\newblock Image{N}et classification with deep convolutional neural networks.
\newblock In {\em NIPS}, 2012.

\bibitem{lang_1997}
P.~Lang, M.~Bradley, and B.~Cuthbert.
\newblock International {A}ffective {P}icture {S}ystem ({IAPS}): {T}echnical
  manual and affective ratings.
\newblock Technical report, NIMH CSEA, 1997.

\bibitem{lecun_1998}
Y.~LeCun, L.~Bottou, Y.~Bengio, and P.~Haffner.
\newblock Gradient-based learning applied to document recognition.
\newblock In {\em Proc. of the IEEE}, 1998.

\bibitem{machajdik_2010}
J.~Machajdik and A.~Hanbury.
\newblock Affective image classification using features inspired by psychology
  and art theory.
\newblock In {\em ACM MM}, 2010.

\bibitem{mcduffpredicting}
D.~McDuff, R.~Kaliouby, J.~Cohn, and R.~Picard.
\newblock Predicting ad liking and purchase intent: Large-scale analysis of
  facial responses to ads.

\bibitem{oquab_2014}
M.~Oquab, L.~Bottou, I.~Laptev, and J.~Sivic.
\newblock Learning and transferring mid-level image representations using
  convolutional neural networks.
\newblock In {\em Computer Vision and Pattern Recognition (CVPR), 2014 IEEE
  Conference on}, pages 1717--1724. IEEE, 2014.

\bibitem{peng_2015}
K.-C. Peng, T.~Chen, A.~Sadovnik, and A.~Gallagher.
\newblock A mixed bag of emotions: {M}odel, predict, and transfer emotion
  distributions.
\newblock In {\em CVPR}, 2015.

\bibitem{plutchik_1980}
R.~Plutchik.
\newblock {\em Emotion: {A} Psychoevolutionary Synthesis}.
\newblock Harper \& Row, 1980.

\bibitem{razavian_2014}
A.~S. Razavian, H.~Azizpour, J.~Sullivan, and S.~Carlsson.
\newblock {CNN} features off-the-shelf: {A}n astounding baseline for
  recognition.
\newblock In {\em Computer Vision and Pattern Recognition Workshops (CVPRW),
  2014 IEEE Conference on}, pages 512--519. IEEE, 2014.

\bibitem{salvador_2015}
A.~Salvador, M.~Zeppelzauer, D.~Manchon-Vizuete, A.~Calafell, and X.~Giro-i
  Nieto.
\newblock Cultural event recognition with visual convnets and temporal models.
\newblock In {\em Computer Vision and Pattern Recognition Workshops (CVPRW),
  2015 IEEE Conference on}. IEEE, 2015.

\bibitem{siersdorfer_2010}
S.~Siersdorfer, E.~Minack, F.~Deng, and J.~Hare.
\newblock Analyzing and predicting sentiment of images on the social web.
\newblock In {\em Proceedings of the international conference on Multimedia},
  pages 715--718. ACM, 2010.

\bibitem{szegedy_2014_googlenet}
C.~Szegedy, W.~Liu, Y.~Jia, P.~Sermanet, S.~Reed, D.~Anguelov, D.~Erhan,
  V.~Vanhoucke, and A.~Rabinovich.
\newblock Going deeper with convolutions.
\newblock {\em arXiv preprint arXiv:1409.4842}, 2014.

\bibitem{szegedy_2014}
C.~Szegedy, W.~Zaremba, I.~Sutskever, J.~Bruna, D.~Erhan, I.~Goodfellow, and
  R.~Fergus.
\newblock Intriguing properties of neural networks.
\newblock In {\em ICLR}, 2014.

\bibitem{tang_2013}
Y.~Tang.
\newblock Deep learning using linear support vector machines.
\newblock In {\em ICML Workshop on Challenges in Representation Learning},
  2013.

\bibitem{xu2014visual}
C.~Xu, S.~Cetintas, K.-C. Lee, and L.-J. Li.
\newblock Visual sentiment prediction with deep convolutional neural networks.
\newblock {\em arXiv preprint arXiv:1411.5731}, 2014.

\bibitem{yanulevskaya_2008}
V.~Yanulevskaya, J.~van Gemert, K.~Roth, A.~Herbold, N.~Sebe, and J.~M.
  Geusebroek.
\newblock Emotional valence categorization using holistic image features.
\newblock In {\em ICIP}, 2008.

\bibitem{you_2015}
Q.~You, J.~Luo, H.~Jin, and J.~Yang.
\newblock Robust image sentiment analysis using progressively trained and
  domain transferred deep networks.
\newblock In {\em The Twenty-Ninth AAAI Conference on Artificial Intelligence
  (AAAI)}, 2015.

\bibitem{zeiler_2014}
M.~D. Zeiler and R.~Fergus.
\newblock Visualizing and understanding convolutional networks.
\newblock In {\em Computer Vision--ECCV 2014}, pages 818--833. Springer, 2014.

\bibitem{zhou2015object}
B.~Zhou, A.~Khosla, A.~Lapedriza, A.~Oliva, and A.~Torralba.
\newblock Object detectors emerge in deep scene cnns.
\newblock 2015.

\end{thebibliography}


\end{document}